# Nice guys don't always finish last: succeeding in hierarchical organizations

Doron Klunover[*]

September 2020

Abstract

What are the chances of an ethical individual rising through the ranks of a political party or a corporation in the presence of unethical peers? To answer this question, I consider a four-player two-stage elimination tournament, in which players are partitioned into those willing to be involved in sabotage behavior and those who are not. I show that, under certain conditions, the latter are more likely to win the tournament.

---

[*] Department of Economics, Ariel University, 40700 Ariel, Israel. E-mail address: doronkl@ariel.ac.il

1. Introduction

It has been acknowledged that office politics is present in almost every workplace and that it usually decreases the efficiency of promotion tournaments among peers dramatically (Carpenter et al., 2010). Nevertheless, it is evident that some individuals avoid such behavior, which is sometimes viewed as unethical, even when it is unobservable and beneficial to them (Charness et al., 2014).[1] Such choices are perhaps related to their intrinsic values. For obvious reasons and for a given level of ability, it is socially desirable that this type of individual is able to get ahead in an organization. But is that usually the outcome? What are the chances of an ethical politician becoming the leader of her party or of an ethical employee moving up the corporate ladder? In contrast to the conventional wisdom, in what follows I will present a theoretical model which shows that their chances are not necessarily less than their unethical competitors. Recent empirical evidence which shows that people with disagreeable personalities do not have an advantage in pursuing power at work may support this result (Anderson et al., 2020)

I consider a two-stage elimination tournament with four risk-neutral contestants, in which two hawks (unethical individuals) and two doves

---

[1] In the experiment conducted by Charness et al. (2014), only around 50% of the participants who had the option of sabotaging others chose to do so, even though it was the equilibrium strategy.

(ethical individuals) compete in pairwise matches. In the first stage of the tournament, each hawk competes against a dove. In the second stage, the winners compete against each other for a single prize allocated to the single winner.[2] In a given match, a hawk chooses productive effort and sabotage (or distractive) effort deducted from the productive effort of her rival, while a dove chooses only productive effort. In particular, in a given match, productive effort increases the player's own chances of winning, while sabotage decreases those of her rival. The terminology of Lazear (1989) who defines the contestant who is less efficient at sabotage as a dove and the one who is more efficient as a hawk is therefore (partially) adopted. More specifically, he defines as a "saint" the contestant whose sabotage effort is sufficiently costly that, under certain conditions, she may choose not to sabotage her rival in equilibrium or may even help her by choosing a negative level of sabotage effort. Note, however, that unlike in Lazear (1989), in this paper it is assumed that the dove's choice not to sabotage is non-strategic and is attributed to intrinsic ethical or moral values. Specifically, the dove's choice

---

[2] The concept of an elimination tournament was used in the seminal paper of Rosen (1986) to describe a similar game with *n* stages without sabotage. It is also used to describe multi-stage contests, in which all players compete against each other, but only the winners survive to the next stage (see, for instance, Altmann et al., 2012 and Fu and Lu, 2012).

not to sabotage is assumed to be exogenously, rather than endogenously, determined.[3]

In the elimination tournament, either all matches are modeled as a contest with a probit contest success function (CSF) (usually defined in the literature as a tournament), which is the canonical model for studying incentive schemes in promotion tournaments (see Chen, 2003, and Münster, 2007), or as a Tullock contest (Tullock, 1980), which is widely used to model lobbying contests (see Konrad 2000; Amegashie, 2012). The main difference between these two types of contests is the way in which noise and effort interact to determine the players' performance.[4]

---

[3] Note that in a rent-seeking contest, neither type of effort is socially desirable (see Congleton et al., 2008). However, sabotage effort may still be viewed as less ethical than standard rent-seeking effort.

[4] In particular, effort in the Tullock contest is multiplied by noise in order to determine performance, while in the tournament-type of contest, noise is added to effort to determine performance. For a review of the literature on sabotage, see Chowdhury and Gürtler (2015) and for a review of contests see Fu and Wu (2019). Note that although Amegashie and Runkel (2007) consider a two-stage elimination tournament with sabotage, they use the "all-pay auction" CSF and only prior to the tournament allow all contestants to sabotage potential rivals who participate in the parallel contest. They build on Groh et al. (2012) who study optimal seedings in a similar game without sabotaging. For the Tullock elimination tournament without sabotage, see Stracke (2013) and Cohen et al. (2018).

It is shown that if the cost function is quasi-linear in productive effort (an assumption that has been considered in the literature; see, for instance, Konrad 2009; pp. 115-118), then in an interior subgame perfect equilibrium (SPE), the probability that a dove wins the prize is greater than half, or if there is only one dove and three hawks or one hawk and three doves, a dove has a higher chance of winning the tournament than a hawk. The main intuition behind these results is that a dove who reaches the finals will be participating in a less intensive competition and therefore will have a larger expected net payoff. Winning in the first stage is therefore more worthwhile for a dove and therefore motivates her to invest greater effort. Although the model is quite simple and therefore cannot fully explain any specific real-world situation, the result is nonetheless grounds for optimism since unobservable ethical behavior can be rewarded. Furthermore, the result may help in understanding why ethical behavior has not become extinct (even though standard evolutionary methodology is not used).

Furthermore, in the case that each match in the elimination tournament is modeled as a Tullock contest, it is shown that a sufficiently large prize insures the existence of a unique interior SPE in pure strategies. In the case where each match is modeled as a tournament, I discuss existence but do not provide a formal proof. Nevertheless, the main result is demonstrated by two examples—one for each type—in which a unique interior SPE exists in pure strategies.

The paper proceeds as follows. Section 2 presents the model. Section 3 presents the main result for contestants' probabilities of winning in an interior SPE. Section 4 discusses existence and section 5 provides examples for the main result. Section 6 concludes.

2. The model

Four players compete in a two-stage elimination tournament. In particular, in the first stage of the tournament, each player $i\epsilon\{1,2,3,4\}$ participates in one contest against a rival $j\epsilon\{1,2,3,4\}$, $j\neq i$. In the second stage, the two winners compete against each other for a prize $v$, which is allocated to the winner. There are two types of players: a Hawk (H) and a Dove (D). In each stage $k\epsilon\{1,2\}$, each player $i$ of type D only invests in "productive" effort $x_i^k \geq 0$ with a unit cost[5], where player $i$ of type H also invests in sabotage effort $s_i^k \geq 0$ with the convex cost function $c(s_i^k)$, where $c(0)=0$, $c'(0)=0$, $c'>0$ and $c''>0$. For any two players $i$ and $j$, $i\neq j$ competing in a contest held in stage $k\epsilon\{1,2\}$, the effective effort of player $i$ is $b_i^k \equiv \max[0, x_i^k - s_j^k]$, while $s_j^k \equiv 0$ if player $j$ is a dove. There are two players of type D and two of type H. In the first stage of the tournament, each player of type D competes against a player of type H.

In a contest held in stage $k\epsilon\{1,2\}$ between players $i$ and $j$, the probability of winning for player $i$ is $p_i^k \equiv p(b_i^k, b_j^k)$ for $i\neq j$. It is assumed that $p_i^k$ is twice differentiable, increasing in $b_i^k$ and decreasing in $b_j^k$ in the interior of its domain. Later on, $p_i^k$ is specified in more detail.

The reminder of the paper focus on pure strategies.

---

[5] Namely, the marginal cost of $x_i^k$ is 1.

3. Main Results

In this section, I characterize an interior SPE, in which $x_i^k>0$ for all $i$ and $s_i^k>0$ for all $i$ of type H for all $k$, and show that, in such an equilibrium, the probability that a dove wins the tournament is greater than half. It is further noted that the result extends to a tournament with one dove and three hawks or one hawk and three doves in the sense that a dove's probability of winning the elimination tournament is larger than that of a hawk. The existence of such an equilibrium is discussed in the following section, after examples are provided to demonstrate the results.

As is usually the practice, the problem is solved by applying backward induction. Note that the contest held in stage 2 is independent of the effort invested in stage 1, and therefore a player $i$ of type H who won the first contest solves:

(1) $\max\limits_{x_i^2, s_i^2} p_i^2 v - c(s_i^2) - x_i^2,$

where a player $i$ of type D solves:

(2) $\max\limits_{x_i^2} p_i^2 v - x_i^2.$

Therefore, the first order conditions (F.O.Cs) for an interior solution in the contest held in stage 2 are as follows:

If both contestants are type D then:

(3) $\frac{\partial p_i^2}{\partial b_i^2} v = 1$

for each contestant $i$.

In the case that both contestants are type H, an interior solution must also satisfy, for each contestant $i \neq j$ and in addition to (3), the condition:

(4) $-\frac{\partial p_i^2}{\partial b_j^2} v = c'(s_i^2)$.

If type H competes against type D, then an interior solution satisfies three conditions: (3) for the type D player, and (3) and (4) for the type H player.

Note that $\frac{\partial p_i^k}{\partial b_i^k} = -\frac{\partial p_j^k}{\partial b_i^k}$ since $p^k(b_i^k, b_j^k) = 1 - p^k(b_j^k, b_i^k)$. Therefore, in a contest held in stage 2 between player $i$ and $j$, condition (3) of player $i$ and condition (4) of player $j$ imply that in an interior solution where at least one of the players is of type H:

(5) $s^2 = c'^{-1}(1)$.

Note that (5) implies that, in an interior solution, $s^2$ is independent of $v$.[6] This in turn implies that an interior solution in which $s^2>0$ may not exist when $v$ is small, since in that case the expected net payoff of a hawk in stage 2 can be negative when $s^2$ satisfies (5). This is further discussed in the next section.

The following lemma determines the contestants' productive effort in an interior symmetric solution in a contest held in stage 2:

**Lemma 1** *Assume that the contest held in stage 2 between player i and player j has an interior symmetric Nash equilibrium in pure strategies, in which $b_i^2=b_j^2$, and let the contestant's effort in a contest held between two doves be $b^*>0$. Then, $b^*+c'^{-1}(1)$ is the productive effort in a contest held between two hawks, and $(b^*, b^*+ c'^{-1}(1))$ is the productive effort pair for a hawk and a dove, respectively, in a contest between them.*

---

[6] This has already been noted by Konrad (2009, pp.115-118).

All proofs appear in the appendix. It follows that in a contest held in stage 2, the expected net payoff of contestant $i$ (not including the sunk cost invested in stage 1) in an interior symmetric solution is:

$$(6)\ E\pi_i^{2^*} = \begin{cases} \frac{v}{2} - b^*, & \text{if both players are type } D \\ \frac{v}{2} - (b^* + c'^{-1}(1) + c(c'^{-1}(1))) & \text{if both players are type } H \\ \frac{v}{2} - (b^* + c'^{-1}(1)), & \text{if only player } i \text{ is of type } D \\ \frac{v}{2} - (b^* + c(c'^{-1}(1))), & \text{if only player } i \text{ is of type } H \end{cases}.$$

Therefore, assuming that the contest held in stage 2 has a unique interior symmetric solution in which $b^2=b^*$, then in a contest held in stage 1 player $i$ of type H solves:

$$(7)\ \max_{x_i^1, s_i^1} p_i^1 \left( \frac{v}{2} - (b^* + c(c'^{-1}(1)) + p_H^1 c'^{-1}(1)) \right) - c(s_i^1) - x_i^1,$$

where $p_H^1$ is the probability that the hawk wins in the parallel contest held in stage 1.

Player $j$ of type D solves:

$$(8)\ \max_{x_j^1} p_j^1 \left( \frac{v}{2} - (b^* + p_H^1 c'^{-1}(1)) \right) - x_j^1.$$

Note that for players $i$ and $j$, $p_H^1$ is taken as given.

Let $A = \frac{v}{2} - (b^* + c(c'^{-1}(1)) + p_H^1 c'^{-1}(1))$ and $B = \frac{v}{2} - (b^* + p_H^1 c'^{-1}(1))$.

The F.O.Cs for an interior solution in the contest held in stage 1 between player $i$ of type H and player $j$ of type D are:

$$(9)\ \frac{\partial p_i^1}{\partial b_i^1} A = 1,$$

$$(10)\ -\frac{\partial p_i^1}{\partial b_j^1} A = c'(s_i^1)$$

and

(11) $\frac{\partial p_j^1}{\partial b_j^1} B = 1$.

To obtain the main result, in the remainder of the analysis it is assumed that in all contests held in the elimination tournament, $p_i^k$ takes one of the following two commonly used CSFs:

a. The Tullock CSF which is formally defined as:

(12) $p_i^k = \begin{cases} \frac{(b_i^k)^r}{(b_i^k)^r+(b_j^k)^r}, & if \ \sum_{t\in\{i,j\}} b_t^k > 0 \\ \frac{1}{2} & otherwise \end{cases}$,

where $r \leq 1$.

This CSF that has been axiomatized by Skaperdas (1996) is considered to be one of the canonical models for analyzing lobbying, sport tournaments, conflicts, etc. (see Corchón and Serena, 2018).

b. The probit CSF (Dixit, 1987) which is usually defined as the "tournament form".

More precisely, let player *i*'s performance in stage *k* be: $y_i^k = f(b_i^k) + \varepsilon_i^k$ for all $i \in \{1,2,3,4\}$ and $k \in \{1,2\}$, where *f* is increasing and concave and $\varepsilon_i^k$ is a random variable distributed symmetrically around zero. In a contest held in stage *k* between two players *i* and *j*, player *i* wins if $y_i^k > y_j^k$. Let *G* be the cumulative distribution function of the variable $\varepsilon_j^k - \varepsilon_i^k$ ($\varepsilon_i^k$ and $\varepsilon_j^k$ are i.i.d). Then:

(13) $p_i^k = G(f(b_i^k) - f(b_j^k))$.

This CSF is widely used to study incentive schemes, and in particular it was used in Lazear (1989), a seminal paper that studied the optimal structure of prizes in promotion tournaments, in which workers also invest effort in sabotage.

It is well known that the main difference between (12) and (13) is that in (12), noise is multiplied by effort to determine player's performance (i.e., $y_i^k = b_i^k \varepsilon_i^k$, where $\varepsilon_i^k$ has the inverse exponential distribution; see Jia, 2008) while in (13), as shown above, noise is added to effort in order to determine performance. In both (12) and (13), the player with the highest performance wins.

I can now state the main result:

**Proposition 1** *Consider an elimination tournament in which $p_i^k$ is defined by (12) or (13) for all k. In an interior SPE in pure strategies, in which $s_i^k \in (0, x_j^k)$ when i is a hawk and $x_i^k > 0$ for all i for all k, the probability that a dove wins v is greater than half.*

The main intuition behind this result is that a contestant in a one-shot contest with a common value prize has a probability of winning of half regardless of her type. However, for a given rival, a contestant of type D invests less effort than a contestant of type H and therefore her expected net payoff is larger. Given that the contest held in stage 2 is equivalent to a one-shot contest with a common value prize, this implies that a dove has a stronger incentive to win the first match than a hawk, and therefore her effective effort is larger. Note that this also implies that if in the elimination tournament there are three hawks and one dove or three doves and one hawk (rather than two hawks and two doves), then a dove has a higher probability of winning the tournament than a hawk.[7]

4. Existence

In this section, I derive sufficient conditions for there to exist a unique interior SPE in pure strategies in the elimination tournament. In particular, Proposition 2 shows that there exists a unique interior SPE in pure strategies when the CSF takes the Tullock form and $v$ is sufficiently large. I also discuss existence for the case in which the CSF is defined by (13).

**Proposition 2** *If $v$ is sufficiently large, and the CSF is defined by (12) in all contests held in the elimination tournament, then there exists a unique interior SPE in pure strategies in the elimination tournament, in which $s_i^k \epsilon (0, x_j^k)$ if i is a hawk, and $x_i^k > 0$ for all i for all $k \epsilon \{1,2\}$.*

The main idea of the proof is that, in any given match and given her rival's equilibrium effort, a hawk's choice set is a compact space and therefore has a global maximum, and when $v$ is sufficiently large that maximum is interior. In the next section, this result is demonstrated by an example, in which $v$ is "reasonably large". Furthermore, the proof of Proposition 2 builds on well-known results for existence of a unique interior equilibrium in pure strategies in the standard (namely, without sabotage) one-shot two-player Tullock contest with either symmetric players (see Perez-Castrillo and Verdier, 1992) or asymmetric players (see Nti, 1999). However, in the tournament-type contest (i.e., with the CSF in (13)), the existence of a unique interior Nash equilibrium in the standard one-shot contest is less obvious. In particular, it has been acknowledged that there must be sufficient dispersion of noise to achieve such an equilibrium (see Lazear and Rosen, 1981; Nalebuff and Stiglitz, 1983; Krishna and Morgan, 1998; Chen, 2003; Drugov and Ryvkin, 2018). Specific conditions depend on the structure of $f$ and the cost function and usually apply to symmetric contests (which does not include the contest held in stage 1). Although important, I abstract from this analysis here. Nevertheless, in order to demonstrate the result in Proposition 1, an example with a CSF defined by (13), in which there exists a unique interior SPE in pure strategies, is presented below. More generally, if there exists an interior equilibrium in the standard one-shot symmetric and asymmetric contests in which the CSF is defined by (13), then a large prize should be sufficient to insure the existence of such an equilibrium in the presence of sabotage.

---

[7] Note that by the reasoning above, in both cases, a dove that competes against a hawk in a contest held in the first stage wins this match with a probability greater than half, while in any other given match held in the tournament both participants have the same winning probability (i.e., a probability of half).

5. Examples

The following examples demonstrate the results: example 1 for the Tullock contest and example 2 for the tournament form. In both, the solution admits the unique interior SPE in pure strategies.

**Example 1** Define $p_i^k$ by (12) and let $r=1$. In addition, assume that $c(s_i^k) = \frac{(s_i^k)^3}{12}$ and $v=80$. Then, by conditions (3) and (4), $s^2=2$ and $b^*=20$ and therefore:

(14) $A = 19\frac{1}{3} - 2p_H^1$

and

(15) $B = 20 - 2p_H^1$.

Let player $i$ be of type H and player $j$ be of type D. Substituting (14) and (15) into (A.3) yields:

(16) $\dfrac{b_i^1}{b_j^1} = \dfrac{19\frac{1}{3} - 2p_H^1}{20 - 2p_H^1} = \dfrac{19\frac{1}{3} - 2\frac{d_i^1}{d_j^1 + d_i^1}}{20 - 2\frac{d_i^1}{d_j^1 + d_i^1}}$,

which implies that, $b_j^1 \approx 1.147 b_i^1$ and therefore:

(17) $p_i^1 \equiv p_H^1 \approx 0.466$.

Substituting (17) into (14) and (15) results in $A \approx 18.4$ and $B \approx 19$. It therefore follows from (A.4) that $s^1 = \sqrt{4\dfrac{A}{B}} \approx 1.97$.

Substituting (17) and (15) into (11) yields $b_i^1 \approx 4.13$ and $b_j^1 \approx 4.73$. Therefore:

(18) $p_i^1 A - c(s^1) - b_i^1 \approx 3.8 > 0$

$(> A - c(b_j^1 + s^1) - \varepsilon \approx -6.8)$

and

(19) $p_j^1 B - (b_j^1 + s^1) \approx 3.46$.

Therefore, according to Proposition 1 and Proposition 2, the probability that a dove wins the tournament in the unique interior SPE in pure strategies is 2*(1/2)*(1-0.466)=0.534.[8]

---

[8] Note that it can be verified that at the solution, $\dfrac{\partial^2 E\pi_i^k}{\partial (s_i^k)^2} < 0$ for all $k \epsilon \{1,2\}$, which is sufficient to conclude that it establish the SPE in pure strategies (see the appendix, footnotes 10, 11 and 13).

**Example 2** Define $p_i{}^k$ by (13) such that $\varepsilon_i{}^k$ and $\varepsilon_j{}^k$ are uniformly distributed over the interval $[-5,5]$[9] and $f(b)=\sqrt{b}$. In addition, assume that $c(s_i^k) = \dfrac{(s_i^k)^3}{27}$ and $v=20$.

Therefore, by (4), $s^2=3$. Furthermore, by condition (3), $\dfrac{\partial p_i^2}{\partial b_i^2}v = g(0)f'v = \dfrac{1}{10}\dfrac{1}{2\sqrt{b_i^2}}20 = 1$, which implies that $b^*=1$. Therefore:

(20) $A = 8 - 3p_H^1$

and

(21) $B = 9 - 3p_H^1$.

Let player $i$ be of type H and player $j$ be of type D. Substituting (20) and (21) into (9) and (11) yields:

(22) $g(\sqrt{b_i^1} - \sqrt{b_j^1})f'(b_i^1)(8 - 3p_H^1)$

$$= \dfrac{2(\sqrt{b_i^1} - \sqrt{b_j^1} + 10)}{200} \dfrac{1}{2\sqrt{b_i^1}}\left(8 - 3\dfrac{(\sqrt{b_i^1} - \sqrt{b_j^1} + 10)^2}{200}\right) = 1$$

and

(23) $g\left(\sqrt{b_i^1} - \sqrt{b_j^1}\right)f'(b_j^1)(9 - 3p_H^1)$

$$= \dfrac{2(\sqrt{b_i^1} - \sqrt{b_j^1} + 10)}{200} \dfrac{1}{2\sqrt{b_j^1}}\left(9 - 3\dfrac{(\sqrt{b_i^1} - \sqrt{b_j^1} + 10)^2}{200}\right) = 1.$$

---

[9] The specification of a unified distribution is commonly used in experimental studies for this type of model (see, for instance, Altmann et al., 2012). Note that this implies that $\varepsilon_j{}^k - \varepsilon_i{}^k$ follows a triangular distribution.

Solving (22) and (23) numerically yields, $\sqrt{b_j^1} \approx 0.373875$ and $\sqrt{b_i^1} \approx 0.324124$. Therefore:

(24) $p_i^1 \equiv p_H^1 \approx \frac{(0.324124 - 0.373875 + 10)^2}{200} = 0.495$.

Substituting (24) into (20) and (21) yields $A = 6.515$ and $B = 7.515$. Substituting into (A.4) yields:

(25) $s^1 = \sqrt{9\frac{A}{B}} \approx 2.79$,

(26) $p_i^1 A - c(s^1) - b_i^1 \approx 2.3$

and

(27) $p_j^1 B - (b_j^1 + s^1) \approx 0.86$.

Note that the second-order conditions for an interior maximum for each player and in any contest are satisfied and it can easily be verified that a contestant $i$ does not benefit from deviating to a corner solution at which $x_i^k \geq 10$ in order to insure that she wins. The first-order necessary conditions for an interior maximum therefore describe the SPE in pure strategies.

6. Conclusion

This paper demonstrates that an ethical individual can become the leader of a political party or the CEO of a corporation. In particular, it is shown that, under some conditions, the players in a four-player elimination tournament who choose not to be involved in sabotage are more likely to win. Although this result is rather surprising and appears to contradict the conventional wisdom, the intuition is quite straightforward and involves standard monetary incentives. Therefore, there is reason to believe that the conclusions carry over to the real world in some circumstances. In other words, it seems that nice guys don't always finish last.

Appendix:

Proof for Lemma 1: Since $p_i^k(b^k, b^k) \equiv p_j^k(b^k, b^k) \equiv p(b^k, b^k)$, then $\frac{\partial p_i^k(b^k,b^k)}{\partial b_i^k} \equiv \frac{\partial p_j^k(b^k,b^k)}{\partial b_j^k}$. Therefore, by condition (3) of players $i$ and $j$, in an interior symmetric solution of a contest held in period 2, $b_i^2 = b_j^2 = b^*$. This implies that, when players $i$ and $j$ of type D compete:

(A.1) $x_i^2 = x_j^2 \equiv b^*$.

When players $i$ and $j$ of type H compete:

(A.2) $x_i^2 = x_j^2 = b^* + c'^{-1}(1)$.

If a player of type H competes against a player of type D, then type D invests $b^* + c'^{-1}(1)$ and the productive effort of type H is $b^*$. QED

Proof of Proposition 1: First, note that if $p_i^2$ is defined by (12) or by (13), then the interior solution of a contest held in stage 2 must be symmetric. Specifically, substituting (12) or (13) into (3) implies that $b_i^2 = b_j^2$ in a contest held in stage 2 between players $i$ and $j$. Therefore, the interior solution of a contest held in stage 2 is defined by Lemma 1.

Furthermore, (9) and (11) imply that in an interior SPE,

(A.3) $\dfrac{\frac{\partial p_j^1}{\partial b_j^1}}{\frac{\partial p_i^1}{\partial b_i^1}} = \frac{A}{B}$.

Specifically, if $p_i^k$ is defined by (12), then $\dfrac{\frac{\partial p_j^1}{\partial b_j^1}}{\frac{\partial p_i^1}{\partial b_i^1}} = \left(\dfrac{b_i^1}{b_j^1}\right)^r$, and if $p_i^k$ is defined by (13), then $\dfrac{\frac{\partial p_j^1}{\partial b_j^1}}{\frac{\partial p_i^1}{\partial b_i^1}} = \dfrac{f'(b_j^1)g(f(b_i^1)-f(b_j^1))}{f'(b_i^1)g(f(b_i^1)-f(b_j^1))} = \dfrac{f'(b_j^1)}{f'(b_i^1)}$. Recall that $f$ is decreasing and therefore either way, $\dfrac{\frac{\partial p_j^1}{\partial b_j^1}}{\frac{\partial p_i^1}{\partial b_i^1}} < 1 \leftrightarrow b_i^1 < b_j^1$. Note that $A < B$ and therefore the RHS of (A.3) is smaller than 1. It follows that if (A.3) is satisfied, then $b_i^1 < b_j^1$ which implies that $p_i^1 < p_j^1$.

Therefore, given that there are two players of type D in the tournament, the probability that this type of player wins $v$ is $2(1/2 p_j^1) = p_j^1 > 1/2$, where $p_j^1$ is the probability that player $j$ of type D wins a contest held in stage 1. QED

Proof for Proposition 2: Define $p_i^k$ by (12). As usual, a backward induction is applied. In particular, I first present an auxiliary lemma that derives a sufficient condition under which there exists a unique interior Nash equilibrium in a contest held in stage 2. Then I proceed to prove the main result.

**Lemma 2** *If $v$ is sufficiently large, then there exists a unique interior Nash equilibrium in a contest held in stage 2.*

*Proof of Lemma* 2: It is well known that the unique solution that solves (3) for both players $i$ and $j$ is $b_i^2 = b_j^2 = rv/4$ (see Perez-Castrillo and Verdier, 1992; Proposition 3). Moreover, in an interior solution, $s^2$ is uniquely defined by (5). Therefore, if the contest held in stage 2 has an interior solution, then it is uniquely defined by Lemma 1 such that $b^*=rv/4$. In particular, if the contest is held between two doves, then it is well known that $b^*=rv/4$ is the unique equilibrium effort (see Perez-Castrillo and Verdier, 1992; Proposition 3). For other cases, it is required to show that Lemma 1 gives the players' best response. This is shown below.

Note that $E\pi_i^2 \leq 0$ at $x_i^2=v$ and therefore without loss of generality I can assume that $x_i^2 \leq v$ for all $i$. Let $x_j^2=rv/4+c'^{-1}(1)$. The choice set of player $i$ of type H is $x_i^2 \times s^2 =[0,v]\times[0,rv/4+c'^{-1}(1)]$, which is a compact set and therefore has a global maximum.[10] Note that this maximum is interior.

To see this, note that $E\pi_i^2\big|_{x_i^2=v} < 0$, and assume first that player $j$ is of type D. Then, $E\pi_i^2\big|_{x_i^2=0} \leq 0$ for all $s^2 < x_j^2$, and $\frac{\partial E\pi_i^2}{\partial s^2}\big|_{s^2=0} = -\frac{\partial p_i^2}{\partial b_j^2} v > 0$ for all $x_i^2 > 0$. Furthermore, if $s^2=x_j^2$, then player $i$ insures her win with an arbitrarily small $x_i^2=\varepsilon$, which implies that $\max E\pi_i^2\big|_{s^2=x_j^2} = v - c\left(\frac{rv}{4}+c'^{-1}(1)\right) - \varepsilon$, while $E\pi_i^2\big|_{(x_i^2,s^2)=\left(\frac{rv}{4},c'^{-1}(1)\right)} = \frac{v(2-r)}{4} - c(c'^{-1}(1))$. Therefore, since $c$ is increasing and convex, $E\pi_i^2\big|_{(x_i^2,s^2)=\left(\frac{rv}{4},c'^{-1}(1)\right)} > \max\left[\max E\pi_i^2\big|_{s^2=x_j^2}, 0\right]$ for sufficiently large $v$.

---

[10] Note that since, by definition, $b_j^k \equiv \max[0,x_j^k-s_j^k]$, the interval $s_i^2 > x_j^2$ is irrelevant. Furthermore, note that when player $j$ is of type D, $E\pi_i^2\big|_{x_j^2>0}$ is continues in $(x_i^2,s^2)$ except at $(x_i^2,s^2)=(0,x_j^2)$, and when player $j$ is of type H, $E\pi_i^2\big|_{x_j^2>0,\, s_j^2>0}$ is continues in $(x_i^2,s_i^2)$ except at $(x_i^2,s_i^2)=(s_j^2,x_j^2)$. However, these two discontinuities are removable since a global maximum is not near them when $v$ is sufficiently large. This is shown below.

Thus, the global maximum of $E\pi_i^2$ is interior. Since at $x_j^2 = rv/4 + c'^{-1}(1)$, $(x_i^2, s^2) = \left(\frac{rv}{4}, c'^{-1}(1)\right)$ is the unique solution to (3) and (4), at which $\frac{\partial^2 E\pi_i^2}{\partial(s_i^2)^2} < 0$, it describes player $i$'s best response to $x_j^2 = rv/4 + c'^{-1}(1)$.[11] Note that since

$-\frac{\partial^2 p_j^2}{\partial (b_j^2)^2}\bigg|_{(x_i^2, s^2, x_j^2) = \left(\frac{rv}{4}, c'^{-1}(1), x\right), x > c'^{-1}(1)} < 0$, (3) describes player $j$'s best response

to $(x_i^2, s^2) = \left(\frac{rv}{4}, c'^{-1}(1)\right)$ when $E\pi_j^2\big|_{x_j^2 = \frac{rv}{4} + c'^{-1}(1)} = \frac{v(2-r)}{4} - c'^{-1}(1) > 0$, which is the case when $v$ is sufficiently large.

---

[11] Specifically, there is another solution to (3) and (4) at which $\frac{\partial^2 E\pi_i^2}{\partial(s_i^2)^2} > 0$ (namely, at this solution the second-order condition for an interior maximum is violated). To see this, note that at $x_j^2 = rv/4 + c'^{-1}(1)$, (3) and (4) imply that $x_i^2(s^2) = c'(s^2)\left(\frac{rv}{4} + c'^{-1}(1) - s^2\right)$. Substituting $x_i^2(s^2)$ in (4) and rearranging terms results in $c'^{-1}(1) + \frac{rv}{4} = s^2 + \frac{rv}{\left((c'(s^2))^{\frac{1+r}{2}} + (c'(s^2))^{\frac{1-r}{2}}\right)^2}$, where the RHS is U-shaped in $s^2$ over the interval $s^2 \in \left[0, c'^{-1}(1) + \frac{rv}{4}\right]$. In particular, since $s^2 + \frac{rv}{\left((c'(s^2))^{\frac{1+r}{2}} + (c'(s^2))^{\frac{1-r}{2}}\right)^2}\bigg|_{s^2 = c'^{-1}(1) + \frac{rv}{4}} > c'^{-1}(1) + \frac{rv}{4}$, and $s^2 + \frac{rv}{\left((c'(s^2))^{\frac{1+r}{2}} + (c'(s^2))^{\frac{1-r}{2}}\right)^2} \to \infty$ as $s^2 \to 0$, there are two roots on the interval $s^2 \in \left(0, c'^{-1}(1) + \frac{rv}{4}\right)$ that solve this equality, one of which is $c'^{-1}(1)$. Note that $\frac{\partial^2 E\pi_i^2}{\partial(s^2)^2}\bigg|_{(x_i^2(s^2), s^2, x_j^2) = (x_i^2(c'^{-1}(1)), c'^{-1}(1), \frac{rv}{4} + c'^{-1}(1))} = \frac{1}{4rv} - c''(c'^{-1}(1))$ is negative for sufficiently large $v$, and therefore, $\frac{\partial^2 E\pi_i^2}{\partial(s_i^2)^2}\bigg|_{(x_i^2, s^2, x_j^2) = (x_i^2(s^2), s^2, \frac{rv}{4} + c'^{-1}(1))}$ must be positive at the other root.

Instead, assume now that player $j$ is of type H and $s_j{}^2=c'^{-1}(1)$, where $v>c'^{-1}(1)$. Note that $E\pi_i^2 \leq 0$ when $x_i{}^2 \leq c'^{-1}(1)$ and $s_i{}^2<x_j{}^2$, and $\left.\frac{\partial E\pi_i^2}{\partial s_i^2}\right|_{s_i^2=0} = -\frac{\partial p_i^2}{\partial b_j^2}v > 0$ for all $x_i^2 > s_j^2$. Furthermore, if $s_i{}^2=x_j{}^2$, then player $i$ insures her win with $x_i{}^2=s_j{}^2+\varepsilon$, which implies that $\max E\pi_i^2\big|_{s_i^2=x_j^2} = v - c\left(\frac{rv}{4} + c'^{-1}(1)\right) - (c'^{-1}(1) + \varepsilon)$, while $E\pi_i^2\big|_{(x_i^2,s_i^2)=\left(\frac{rv}{4}+c'^{-1}(1),c'^{-1}(1)\right)} = \frac{v(2-r)}{4} - c'^{-1}(1) - c(c'^{-1}(1))$. Therefore, since $c$ is increasing and convex, for $v$ sufficiently large, $E\pi_i^2\big|_{(x_i^2,s_i^2)=\left(\frac{rv}{4}+c'^{-1}(1),c'^{-1}(1)\right)} > \max\left[\max E\pi_i^2\big|_{s^2=x_j^2}, 0\right]$ for sufficiently large $v$. This implies that the global maximum of $E\pi_i^2$ is interior and therefore, $(x_i^2, s_i^2) = \left(\frac{rv}{4} + c'^{-1}(1), c'^{-1}(1)\right)$ describes player $i$'s best response to itself (i.e., to: $(x_j^2, s_j^2) = \left(\frac{rv}{4} + c'^{-1}(1), c'^{-1}(1)\right)$).[12] QED

I now proceed to stage 1. Assume that $v$ is sufficiently large such that Lemma 2 holds. Then a contest held in stage 1 is equivalent to a one-shot contest between player $i$ of type H and player $j$ of type D, in which $A$ is contestant's *i*'s evaluation of the prize and *B* is contestant's *j*'s evaluation fo the prize. In what follows, it is assumed that this is the case.

---

[12] Note that $(x_i^2, s_i^2) = \left(\frac{rv}{4} + c'^{-1}(1), c'^{-1}(1)\right)$ is the only solution of (3) and (4) at which $\frac{\partial^2 E\pi_i^2}{\partial (s_i^2)^2} < 0$. In particular, at $x_j^2 = \frac{rv}{4} + c'^{-1}(1)$, (3) and (4) imply that $x_i^2(s_i^2) = c'(s_i^2)\left(\frac{rv}{4} + c'^{-1}(1) - s_i^2\right) + s_j^2$ and $c'^{-1}(1) + \frac{rv}{4} = s_i^2 + \frac{rv}{\left((c'(s_i^2))^{\frac{1+r}{2}} + (c'(s_i^2))^{\frac{1-r}{2}}\right)^2}$, which implies that $s_i^2$ (and also $\left.\frac{\partial^2 E\pi_i^2}{\partial (s_i^2)^2}\right|_{(x_i^2(s_i^2),s_i^2,x_j^2)}$) is determined independently of $s_j^2$ and therefore the analysis proceeds as in footnote 10.

Note that by (10) and (11) in an interior solution of this contest:

(A.4) $s^1 = c'^{-1}(\frac{A}{B})$.

Furthermore, by Nti (1999, Proposition 3), the unique solution of (9) and (11) is $(b_i^1, b_j^1) = (\frac{rA^{r+1}B^r}{(A^r+B^r)^2}, \frac{rB^{r+1}A^r}{(A^r+B^r)^2})$. Therefore, if an interior solution exists, then it is unique (namely, there can only be one interior solution).

Assume that $x_j^1 = \frac{rB^{r+1}A^r}{(A^r+B^r)^2} + c'^{-1}(\frac{A}{B})$. Then, the choice set of player *i* is $x_i^1 \times s^1 = [0,A] \times [0,x_j^1]$, which is a convex set and therefore has a global maximum.[13] Note that $E\pi_i^1|_{x_i^1=A} < 0$, $E\pi_i^1|_{x_i^1=0} \leq 0$ for all $s^1 < x_j^1$ and $\frac{\partial E\pi_i^1}{\partial s^1}|_{s^1=0} = -\frac{\partial p_i^1}{\partial b_j^1}A > 0$ for all $x_i^1 > 0$. Furthermore, if $s^1 = x_j^1$, then player *i* guarantees her win with an arbitrarily small $x_i^1 = \varepsilon$, which implies that

$$\max E\pi_i^1|_{s^1=x_j^1} = A - c\left(\frac{rB^{r+1}A^r}{(A^r+B^r)^2} + c'^{-1}\left(\frac{A}{B}\right)\right) - \varepsilon, \quad \text{while}$$

$E\pi_i^1|_{(x_i^1,s^1)=\left(\frac{rA^{r+1}B^r}{(A^r+B^r)^2}, c'^{-1}(\frac{A}{B})\right)} = \frac{A^{r+1}}{(A^r+B^r)^2}(A^r + B^r - rB^r) - c(c'^{-1}(\frac{A}{B}))$. Note that when $v \to \infty$, $t \to \frac{v(2-r)}{4}$ $t\epsilon\{A,B\}$ and therefore, since *c* is increasing and convex, for *v* sufficiently large, $E\pi_i^1|_{(x_i^1,s^1)=\left(\frac{rA^{r+1}B^r}{(A^r+B^r)^2}, c'^{-1}(\frac{A}{B})\right)} > \max\left[\max E\pi_i^1|_{s^1=x_j^1}, 0\right]$ for sufficiently large *v*, which implies that the global maximum of $E\pi_i^1$ is interior and therefore $(x_i^1, s^1) = \left(\frac{rA^{r+1}B^r}{(A^r+B^r)^2}, c'^{-1}(\frac{A}{B})\right)$ describes player *i*'s best response to

---

[13] The comment in footnote 9 regarding a contest held in stage 2, in which player *i* is of type D, applies here as well.

$x_j^1 = \frac{rB^{r+1}A^r}{(A^r+B^r)^2} + c'^{-1}(\frac{A}{B})$.[14] Furthermore, since $-\frac{\partial^2 p_j^1}{\partial (b_j^1)^2}\bigg|_{(x_i^1,s^1)=\left(\frac{rA^{r+1}B^r}{(A^r+B^r)^2}, c'^{-1}(\frac{A}{B})\right)} < 0$

for all $x_j^1 > 0$, (11) describes player $j$'s best response when

$E\pi_j^1\bigg|_{\left(x_j^1,x_i^1,s^1\right)=\left(\frac{rA^rB^{r+1}}{(A^r+B^r)^2}+c'^{-1}(\frac{A}{B}),\frac{rA^{r+1}B^r}{(A^r+B^r)^2},c'^{-1}(\frac{A}{B})\right)} = \frac{B^{r+1}}{(A^r+B^r)^2}(A^r+B^r-rA^r) -$

---

[14] Note that at $x_j^1 = \frac{rB^{r+1}A^r}{(A^r+B^r)^2} + c'^{-1}(\frac{A}{B})$, (9) and (10) imply that

$x_i^1(s^1) = c'(s^1)\left(\frac{rB^{r+1}A^r}{(A^r+B^r)^2} + c'^{-1}(\frac{A}{B}) - s^1\right)$. Substituting $x_i^1(s^1)$ into (10) and

rearranging terms yields $\frac{rB^{r+1}A^r}{(A^r+B^r)^2} + c'^{-1}(\frac{A}{B}) = s^1 + \frac{rA}{\left((c'(s^1))^{\frac{1+r}{2}}+(c'(s^1))^{\frac{1-r}{2}}\right)^2}$, in which the RHS is U-shaped in $s^1$ over the interval $s^1 \in \left[0, \frac{rB^{r+1}A^r}{(A^r+B^r)^2} + c'^{-1}(\frac{A}{B})\right]$. In particular, since $s^1 + \frac{rA}{\left((c'(s^1))^{\frac{1+r}{2}}+(c'(s^1))^{\frac{1-r}{2}}\right)^2}\bigg|_{s^1=\frac{rB^{r+1}A^r}{(A^r+B^r)^2}+c'^{-1}(\frac{A}{B})} > \frac{rB^{r+1}A^r}{(A^r+B^r)^2} + c'^{-1}(\frac{A}{B})$ and $s^1 +$

$\frac{rA}{\left((c'(s^1))^{\frac{1+r}{2}}+(c'(s^1))^{\frac{1-r}{2}}\right)^2} \to \infty$ as $s^1 \to 0$, there are two roots on the interval $s^1 \in$

$\left(0, \frac{rB^{r+1}A^r}{(A^r+B^r)^2} + c'^{-1}(\frac{A}{B})\right)$ that solve this equality, where $c'^{-1}(\frac{A}{B})$ is one of them. Note

that when $v \to \infty$

$\frac{\partial^2 E\pi_i^1}{\partial (s^1)^2}\bigg|_{\left(x_i^1(s^1),s^1,x_j^1\right)=\left(x_i^1(c'^{-1}(\frac{A}{B})),c'^{-1}(\frac{A}{B}),\frac{rB^{r+1}A^r}{(A^r+B^r)^2}+c'^{-1}(\frac{A}{B})\right)} = \frac{(A^r+B^r)((1+r)A^r+(1-r)B^r)}{rA^{2r-1}B^2} -$

$c''\left(c'^{-1}(\frac{A}{B})\right) \to \frac{1}{4rv} - c''(c'^{-1}(1))$, which is negative for sufficiently large $v$, and

therefore, $\frac{\partial^2 E\pi_i^1}{\partial (s^1)^2}\bigg|_{\left(x_i^1,s^1,x_j^1\right)=\left(x_i^1(s^1),s^1,\frac{rB^{r+1}A^r}{(A^r+B^r)^2}+c'^{-1}(\frac{A}{B})\right)}$ must be positive at the other root.

$c'^{-1}(\frac{A}{B}) > 0$, which is the case when $v$ is sufficiently large (since then $t \to \frac{v(2-R)}{4}$, $t\epsilon\{A,B\}$).[15] QED

References

Altmann, S., Falk, A. and Wibral, M., 2012. Promotions and incentives: The case of multistage elimination tournaments. *Journal of Labor Economics*, 30(1), pp.149-174.

Anderson, C., Sharps, D.L., Soto, C.J. and John, O.P., 2020. People with disagreeable personalities (selfish, combative, and manipulative) do not have an advantage in pursuing power at work. *Proceedings of the National Academy of Sciences*.

Amegashie, J.A. and Runkel, M., 2007. Sabotaging potential rivals. *Social Choice and Welfare*, 28(1), pp.143-162.

Amegashie, J.A., 2012. Productive versus destructive efforts in contests. *European Journal of Political Economy*, 28(4), pp.461-468.

Carpenter, J., Matthews, P.H. and Schirm, J., 2010. Tournaments and office politics: Evidence from a real effort experiment. *American Economic Review*, 100(1), pp.504-17.

Charness, G., Masclet, D. and Villeval, M.C., 2014. The dark side of competition for status. *Management Science*, 60(1), pp.38-55.

---

[15] Note that with further specification of $p_i^k$ and $c(s^k)$, $x_i^1$ and $x_j^1$ can be obtained for instance by dividing (11) by (9) to calculate, $p_H^1$, and then substituting $p_H^1$ into $A$ and $B$ to obtain: $(b_i^1, b_j^1) = (\frac{rA^{r+1}B^r}{(A^r+B^r)^2}, \frac{rB^{r+1}A^r}{(A^r+B^r)^2})$. This is demonstrated in Example 1.

Chen, K.P., 2003. Sabotage in promotion tournaments. *Journal of Law, Economics, and Organization*, 19(1), pp.119-140.

Chowdhury, S.M. and Gürtler, O., 2015. Sabotage in contests: a survey. *Public Choice*, *164*(1-2), pp.135-155.

Cohen, N., Maor, G. and Sela, A., 2018. Two-stage elimination contests with optimal head . *Review of Economic Design*, *22*(3-4), pp.177-192.

Congleton, R.D., Hillman, A.L. and Konrad, K.A. eds., 2008 40 .Years of Research on Rent Seeking 2: Applications: Rent Seeking in Practice) Vol. 2). Springer Science & Business Media.

Corchón, L.C., Serena, M., 2018. Contest theory. In: *Handbook of Game Theory and Industrial Organization, Volume II: Applications*, 125.

Dixit, A., 1987. Strategic behavior in contests. *The American Economic Review*, pp.891-898.

Drugov, M. and Ryvkin, D., 2018. Tournament rewards and heavy tails (No. w0250).

Fu, Q. and Wu, Z., 2019. Contests: Theory and topics. In *Oxford Research Encyclopedia of Economics and Finance*.

Fu, Q. and Lu, J., 2012. The optimal multi-stage contest. *Economic Theory*, *51*(2), pp.351-382.

Groh, C., Moldovanu, B., Sela, A. and Sunde, U., 2012. Optimal seedings in elimination tournaments. *Economic Theory*, 49(1), pp.59-80.

Jia, H., 2008. A stochastic derivation of the ratio form of contest success functions. *Public Choice*, *135*(3-4), pp.125-130.

Konrad, K.A., 2000. Sabotage in rent-seeking contests. *Journal of Law,

*Economics, and Organization*, 16(1), pp.155-165.

Konrad, K.A., 2009. *Strategy and dynamics in contests*. Oxford University Press, Oxford U.K

Krishna, V. and Morgan, J., 1998. The winner-take-all principle in small tournaments. *Advances in applied microeconomics*, 7, pp.61-74.

Lazear, E.P., 1989. Pay equality and industrial politics. *Journal of political economy*, 97(3), pp.561-580.

Lazear, E.P. and Rosen, S., 1981. Rank-order tournaments as optimum labor contracts. *Journal of political Economy*, *89*(5), pp.841-864.

Münster, J., 2007. Selection tournaments, sabotage, and participation. *Journal of Economics & Management Strategy*, 16(4), pp.943-970.

Nalebuff, B.J. and Stiglitz, J.E., 1983. Prizes and incentives: towards a general theory of compensation and competition. *The Bell Journal of Economics*, pp.21-43.

Nti, K.O., 1999. Rent-seeking with asymmetric valuations. *Public Choice*, *98*(3-4), pp.415-430.

Perez-Castrillo, J.D. and Verdier, T., 1992. A general analysis of rent-seeking games. *Public choice*, *73*(3), pp.335-350.

Rosen, S., 1986. Prizes and Incentives in Elimination Tournaments. *The American Economic Review*, pp.701-715.

Skaperdas, S., 1996. Contest success functions. *Economic theory*, 7(2), pp.283-290.

Stracke, R., 2013. Contest design and heterogeneity. *Economics Letters*, 121(1), pp.4-7.

Tullock, G., 1980. Efficient rent seeking. In: Buchanan, J.M., Tollison, R.D.,

Tullock, G. (Eds.), *Towards a Theory of the Rent-seeking Society*. Texas A&M University Press, College Station, TX, pp. 97–112.